# Superconductivity phase diagram of Se-substituted $CeO_{0.5}F_{0.5}Bi(S_{1-x}Se_x)_2$

**Yoshikazu Mizuguchi, Takafumi Hiroi, Osuke Miura**

Tokyo Metropolitan University, 1-1, Minami-osawa, Hachioji 192-0397, Japan

mizugu@tmu.ac.jp

**Abstract**. We investigated the effects of Se substitution on the lattice constants and superconducting properties of $CeO_{0.5}F_{0.5}Bi(S_{1-x}Se_x)_2$. With increasing Se concentration, the $a$ lattice constant increased, while the $c$ lattice constant did not show any significant increase between $x = 0.1$ and $x = 0.5$. Bulk superconductivity was observed in samples with $x = 0.2$–$0.4$, and the superconducting transition temperature was the highest at $x = 0.3$. The obtained superconductivity phase diagram was compared to those of $LaO_{0.5}F_{0.5}Bi(S_{1-x}Se_x)_2$ and $NdO_{0.5}F_{0.5}Bi(S_{1-x}Se_x)_2$.

1. **Introduction**

$BiCh_2$-based (Ch: S, Se) layered superconductors have been drawing considerable attention because of the superconductivity observed in layered structures composed of conducting $BiCh_2$ layers alternating with electrically insulating blocking layers, which resemble the structures of cuprate or Fe-based high-transition-temperature ($T_c$) superconductors [1,2]. Since the discovery of superconductivity in $Bi_4O_4S_3$ [3] and $LaO_{1-x}F_xBiS_2$ [4], several types of $BiS_2$-based superconductors have been discovered [5]. In addition, a $BiSe_2$-based superconductor ($LaO_{0.5}F_{0.5}BiSe_2$) was also synthesized [6]. However, the superconductivity mechanisms of $BiCh_2$-based superconductors have not been clarified yet.

Superconductivity in $BiCh_2$-based materials is essentially induced by electron carrier doping into Bi-6p orbitals; parent compounds, such as $LaOBiS_2$, are insulators with band gaps [7]. In addition, a recent synchrotron X-ray diffraction study indicated that crystal structure optimization is also required for the appearance of bulk superconductivity in $REOBiCh_2$-type (RE: rare earth) materials [8]. Specifically, bulk superconductivity states can be achieved when both the electron carrier concentration and the crystal structure are optimized. We have systematically studied the relationship between superconductivity and crystal structure by tuning the *chemical pressure*, which is lattice shrinkage (or expansion) not accompanied by carrier doping. For example, in $LaOBiS_2$, the chemical pressure can be tuned by substituting $La^{3+}$ (1.16 Å) with $Ce^{3+}$ (1.14 Å) or $Nd^{3+}$ (1.12 Å) [9] or by substituting $S^{2-}$ (1.84 Å) with $Se^{2-}$ (1.98 Å). In $LaO_{0.5}F_{0.5}Bi(S_{1-x}Se_x)_2$, bulk superconductivity is induced with increasing Se concentration, and the highest $T_c$ is observed when $x = 0.5$ [10]. In $NdO_{0.5}F_{0.5}Bi(S_{1-x}Se_x)_2$, however, $T_c$ decreases with Se substitution [11]. Having considered the fact that these Se substitutions in $REO_{0.5}F_{0.5}Bi(S_{1-x}Se_x)_2$ do not dope electron carriers, clarifying the universal relationship between the superconductivity and the crystal structure of Se-substituted $REO_{0.5}F_{0.5}Bi(S_{1-x}Se_x)_2$ should provide information useful for understanding the superconductivity mechanisms in $BiCh_2$-based superconductors. Herein, we present the effects of Se substitution on the crystal structure (lattice constants) and superconducting properties of $CeO_{0.5}F_{0.5}Bi(S_{1-x}Se_x)_2$.

## 2. Experimental details

Polycrystalline samples of $CeO_{0.5}F_{0.5}Bi(S_{1-x}Se_x)_2$ were prepared using the solid-state-reaction method. Initial powders of $CeO_2$ (99.99%), $CeF_3$ (99.9%), $Ce_2S_3$ (99.9%), $Bi_2Se_3$ (synthesized from Bi (99.999%) and Se (99.999%) grains), Bi, and S (99.99%) with the nominal compositions of the target phases ($x = 0.1$–$0.5$) were mixed, pelletized, and sealed into evacuated quartz tubes. The samples were heated for 15 h at 700 °C. The obtained (reacted) pellets were ground, pelletized, sealed into evacuated quartz tubes, and heated for an additional 15 h at 700 °C. The purities and lattice constants of the samples were examined using powder X-ray diffraction (XRD) with CuKα radiation and the $\theta$–$2\theta$ method. Their superconducting properties were investigated by performing magnetic susceptibility ($\chi$) measurements using a superconducting quantum interference device (SQUID) magnetometer with an applied field of about 5 Oe after zero-field cooling down to 2 K.

## 3. Results and discussion

Figure 1(a) shows the XRD patterns for $CeO_{0.5}F_{0.5}Bi(S_{1-x}Se_x)_2$. The target phase of $CeO_{0.5}F_{0.5}Bi(S_{1-x}Se_x)_2$ was obtained for $x \leq 0.5$. For $x = 0.1$–$0.3$, almost all of the XRD peaks could be indexed using the space group of $P4/nmm$ [2] except for the small peaks due to impure $CeF_3$, which are indicated with circles (●). For $x = 0.4$ and 0.5, peaks resulting from impure $Bi_2Se_3$ are evident. Based on these results, we assume that the solubility limit of S and Se in the $CeO_{0.5}F_{0.5}Bi(S_{1-x}Se_x)_2$ structure exists near $x = 0.5$. The lattice constants were calculated from the peak positions of the (200) and (004) reflections. The calculated values of the $a$ and $c$ lattice constants are plotted in Figs. 1(b) and 1(c), respectively, as functions of the nominal $x$ value. The $a$ lattice constant increases monotonically with increasing Se concentration, while the $c$ lattice constant firstly increases (from $x = 0$ [9] to $x = 0.1$) but does not significantly change between $x = 0.1$ and $x = 0.5$. These lattice constant evolutions suggest that the substituted Se ions occupy in-plane Ch sites (see Fig. 1(d)). This tendency is similar to that observed in $LaO_{0.5}F_{0.5}Bi(S_{1-x}Se_x)_2$ [10]. The anisotropic elongations of the $a$- and $c$-axes should affect the structure of the $BiCh_5$ pyramid, which is an important factor in the superconductivity of $BiCh_2$-based materials [12,13]. To analyze the crystal structure parameters of $CeO_{0.5}F_{0.5}Bi(S_{1-x}Se_x)_2$, further synchrotron XRD, X-ray absorption spectroscopy, and/or single-crystal structural analyses are necessary.

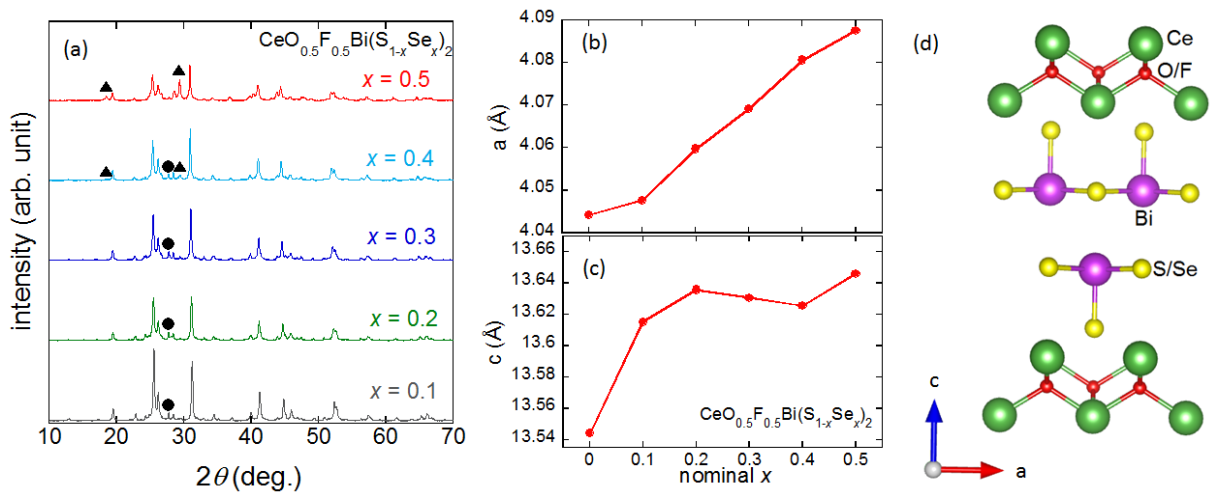

Fig. 1. (a) XRD patterns for $CeO_{0.5}F_{0.5}Bi(S_{1-x}Se_x)_2$. Triangles (▲) and circles (●) indicate peaks resulting from impure $Bi_2Se_3$ and $CeF_3$, respectively. (b) Nominal $x$ dependence of $a$ lattice constant of $CeO_{0.5}F_{0.5}Bi(S_{1-x}Se_x)_2$. (c) Nominal $x$ dependence of $c$ lattice constant of $CeO_{0.5}F_{0.5}Bi(S_{1-x}Se_x)_2$. (d) Schematic image of crystal structure of $CeO_{0.5}F_{0.5}Bi(S_{1-x}Se_x)_2$. Image was drawn using VESTA [14].

Figure 2(a) shows the temperature dependences of the magnetic susceptibilities ($4\pi\chi$) of $CeO_{0.5}F_{0.5}Bi(S_{1-x}Se_x)_2$ with different $x$ values. Superconducting (diamagnetic) signals are not apparent for $x \leq 0.1$. Large diamagnetic signals are evident for $x$ = 0.2–0.4, indicating the emergence of bulk superconductivity states in these samples. For $x$ = 0.5, the superconducting transition is not observable. Using the estimated values of $T_c$, we established a superconductivity phase diagram for $CeO_{0.5}F_{0.5}Bi(S_{1-x}Se_x)_2$ (Fig. 2(b)). The phase diagram exhibits dome-shaped behavior upon Se substitution, which is similar to that in the phase diagram of $LaO_{0.5}F_{0.5}Bi(S_{1-x}Se_x)_2$ [10]. However, the $x$ value at the peak of the $CeO_{0.5}F_{0.5}Bi(S_{1-x}Se_x)_2$ diagram ($x$ = 0.3) is significantly different from that of the $LaO_{0.5}F_{0.5}Bi(S_{1-x}Se_x)_2$ diagram ($x$ = 0.5). Furthermore, the phase diagram for $NdO_{0.5}F_{0.5}Bi(S_{1-x}Se_x)_2$ does not exhibit dome-shaped behavior; instead, $T_c$ monotonically decreases with increasing $x$ [11]. To compare these $REO_{0.5}F_{0.5}Bi(S_{1-x}Se_x)_2$ superconductivity phase diagrams, all three are plotted in Fig. 3.

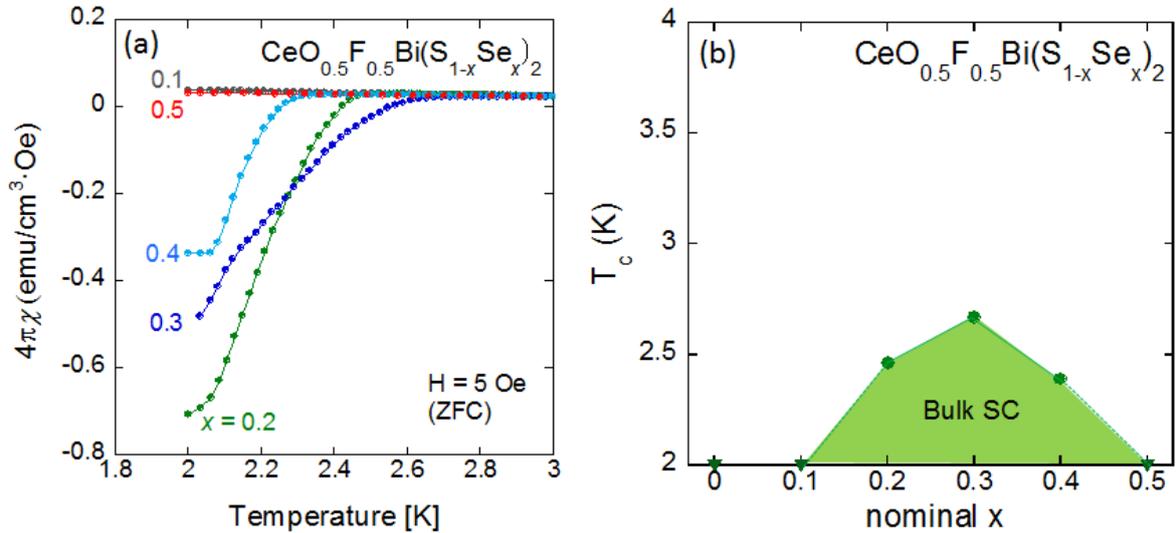

Fig. 2. (a) Temperature dependences of magnetic susceptibilities ($4\pi\chi$) for $CeO_{0.5}F_{0.5}Bi(S_{1-x}Se_x)_2$ with different $x$ values. (b) Superconductivity phase diagram for $CeO_{0.5}F_{0.5}Bi(S_{1-x}Se_x)_2$. Bulk SC denotes region showing bulk superconductivity. Inverted triangles (▼) indicate samples showing no superconductivity above 2 K (measurement limit of our SQUID system).

Figure 3 presents the phase diagrams for $REO_{0.5}F_{0.5}Bi(S_{1-x}Se_x)_2$ (RE = La, Ce, and Nd) as functions of the nominal $x$ value. For $LaO_{0.5}F_{0.5}Bi(S_{1-x}Se_x)_2$, bulk superconductivity is apparent for $x \geq 0.2$, while filamentary (weak) superconducting signals are observable at $x$ = 0 and 0.1. $T_c$ increases as $x$ increases to 0.5, and it decreases above $x$ = 0.5. For $CeO_{0.5}F_{0.5}Bi(S_{1-x}Se_x)_2$, bulk superconductivity is apparent at $x$ = 0.2, and the highest $T_c$ occurs at $x$ = 0.3. With further Se substitution, superconductivity is suppressed. For $NdO_{0.5}F_{0.5}Bi(S_{1-x}Se_x)_2$, the $T_c$ value of $NdO_{0.5}F_{0.5}BiS_2$ ($x$ = 0) is the highest ($T_c \sim 5$ K) among all of the $REO_{0.5}F_{0.5}Bi(S_{1-x}Se_x)_2$ samples in this study, but it decreases with Se substitution. Substituting S with Se in $NdO_{0.5}F_{0.5}Bi(S_{1-x}Se_x)_2$ degrades its superconductivity. These differences in the effects of Se substitution on superconductivity should be related to the differences in the REO blocking layers in these materials, which are related to their chemical pressures. As mentioned in the introduction, the ionic radii of $RE^{3+}$ are different: assuming a coordination number of 8, $La^{3+}$, $Ce^{3+}$, and $Nd^{3+}$ have ionic radii of 1.16 Å, 1.14 Å, and 1.12 Å, respectively. The peak superconductivities for RE = La, Ce, and Nd occur at $x$ = 0.5, 0.3, and 0 (or less than 0), respectively. Therefore, the $x$ values at the peaks seem to be correlated with the sizes of the blocking layers in these materials.

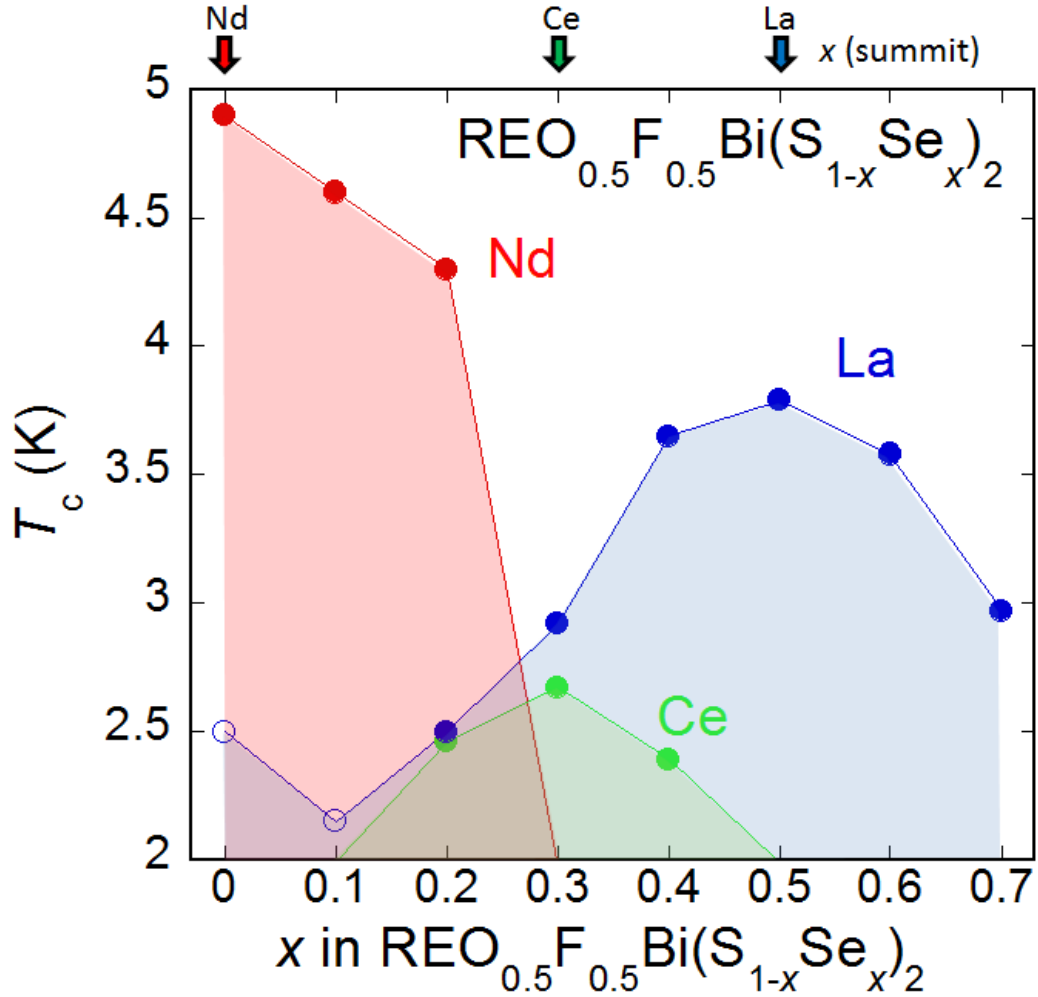

Fig. 3. Superconductivity phase diagrams for $REO_{0.5}F_{0.5}Bi(S_{1-x}Se_x)_2$ (RE = La, Ce, and Nd). Open blue circles for $LaO_{0.5}F_{0.5}Bi(S_{1-x}Se_x)_2$ indicate samples showing filamentary superconductivity.

## 4. Conclusion

We studied the effects of Se substitution on the crystal structure and superconductivity of $CeO_{0.5}F_{0.5}Bi(S_{1-x}Se_x)_2$. With increasing Se concentration, the $a$ lattice constant increased, while the $c$ lattice constant did not show any significant increase between $x$ = 0.1 and $x$ = 0.5. Bulk superconductivity was observed for the samples with $x$ = 0.2–0.4, and the highest $T_c$ value was attained at $x$ = 0.3. Comparing the phase diagrams for $REO_{0.5}F_{0.5}Bi(S_{1-x}Se_x)_2$, we suggested that the $x$ values yielding the peak superconductivities were related to the sizes of their REO blocking layers ($RE^{3+}$ ionic radii).

**Acknowledgements**

This work was partly supported by Grant-in-Aid for Young Scientist (A) (25707031), Grant-in-Aid for challenging Exploratory Research (26600077), and Grant-in-Aid for Scientific Research on Innovative Areas (15H05886).

**References**

[1]   J. B. Bednorz, and K. Müller 1986 Z. Physik B Condens. Matter **64** 189-193.

[2]   Y. Kamihara *et al.* 2008 J. Am. Chem. Soc. **130** 3296-3297.

[3]   Y. Mizuguchi *et al.* 2012 Phys. Rev. B **86** 220510(1-5).

[4]   Y. Mizuguchi *et al.* 2012 J. Phys. Soc. Jpn. **81** 114725(1-5).

[5]   Y. Mizuguchi 2015 J. Phys. Chem. Solid **84** 34-48.

[6]   A.K. Maziopa *et al.* 2014 J. Phys.: Condens. Matter **26** 215702(1-5).

[7]   H. Usui, K. Suzuki, and K. Kuroki 2012 Phys. Rev. B **86** 220501(1-5).

[8]   Y. Mizuguchi *et al.* 2015 Sci. Rep. **5** 14968(1-8).

[9]   J. Kajitani, *et al.* 2015 J. Phys. Soc. Jpn. **84** 044712(1-6).

[10]  T. Hiroi *et al.* 2015 J. Phys. Soc. Jpn **84** 024723(1-4).

[11]  T. Hiroi *et al.* 2014 J Supercond. Nov. Magn. **28** 1149-1153.

[12]  T. Sugimoto *et al.* 2014 Phys. Rev. B **89** 201117(1-5).

[13]  E. Paris *et al.* 2014 J. Phys.: Condens. Matter **26** 435701(1-6).

[14]  K. Momma, and F. Izumi 2008 J. Appl. Crystallogr. **41** 653-658.